\documentclass[prb,aps,twocolumn,floats,showpacs,amsmath,amssymb]{revtex4-1}
\usepackage{bm,graphicx}
\usepackage{amsmath}
\newcommand{\be}{\begin{equation}}
\newcommand{\ee}{\end{equation}}
\newcommand{\bea}{\begin{eqnarray}}
\newcommand{\eea}{\end{eqnarray}}

\newcommand{\bib}{\bibitem}

\usepackage{graphicx}
\usepackage{color}
\usepackage{epsfig}
\usepackage{epstopdf}
\usepackage{amsmath,amssymb}
\usepackage{hyperref}

\begin{document}

\title{Hofstadter butterfly in the Falicov-Kimball model on some finite 2D lattices}
\author{Subhasree Pradhan}
\email[E-mail: ]{spradhan@phy.iitkgp.ernet.in }

\affiliation{ Department of Physics, Indian Institute of Technology Kharagpur, Kharagpur 721302, India}

\begin{abstract}

Spinless, interacting electrons on a finite size triangular lattice moving in an extremely strong perpendicular magnetic field are studied and compared with the results on a square lattice. Using a Falicov-Kimball model, the effects of the magnetic field, Coulomb correlation and finite system size on their energy spectrum are observed. It is possible to induce a gap in the spectrum by tuning the magnetic field even in the absence of correlation, though extra states appear in the gap due to finite size. An orbital current is calculated for both the square and triangular lattice with and without electron correlation. In the noninteracting limit, the bulk current shows several patterns, while the edge current shows oscillations with magnetic flux. The oscillations persist in the interacting limit for the square lattice but not for the triangular lattice. Using exact diagonalization techniques, the recursive structure of Hofstadter spectrum is examined with strong electronic correlations for different system sizes. Electronic correlation is found to suppress these extra states in the gap in some cases.

\end{abstract}

\date{\today}
\maketitle

\section{Introduction}

The problem of electrons moving in a periodic potential under the influence of an external magnetic field has been rigorously investigated, giving rise to several phenomena such as Quantum Hall effect\cite{QHE1, QHE2}, superconducting flux phases~\cite{Lederer} and the famous Hofstadter butterfly~\cite{Hofstadter}, to name a few. Despite the simplicity of the problem, many aspects of it still remain unresolved due to the paucity of both experimental and exact theoretical results. One can think of the problem in two limiting cases, in one, the lattice potential is much weaker than the cyclotron frequency, the problem reduces to free electron Landau eigenstates. On the other hand, when the strength of the periodic potential is comparable to the cyclotron energy, one can solve the tight-binding Hamiltonian of the Bloch electrons by Peierls substitution in the hopping term, which produces the Hofstadter butterfly \cite{Hofstadter, Langbein}. The energy spectrum is found to depend critically on the ratio $p/q$ ($p$ and $q$ are positive integers, the ratio of the magnetic flux per plaquette to the flux quantum). If $p/q$ is a rational number, each energy band is split into $q$ subbands by the magnetic field. The Hofstadter butterfly reveals a recursive structure for a rational flux and a Cantor set at an irrational flux. The fascinating nature of the butterfly is a characteristic feature of 2D systems. Similar to the case of Landau levels, the dimensionality of the system is of crucial importance for the Hofstadter butterfly. The movement of electrons in the external magnetic field is responsible for the broadening of the Hofstadter bands. They may eventually overlap and smear out the original fractal structure of the energy spectrum (see Fig.1). There are a number of theoretical studies to reveal this fractal spectrum in presence of several parameters such as electron-phonon coupling \cite {polaron, Mona1}, Rashba-spin-orbit coupling \cite{rashba} and disorder \cite{disorder}. The main challenge in realizing this recursive structure is the requirement of extremely high magnetic field. There are recent indications of this kind of structure in some artificial superlattices by enhancing the lattice scale to the magnetic length scale~\cite{superlattice, QHC}. On the other hand, remarkable developments of experimental techniques in ultracold gases in recent years have allowed search for novel states of matter which go beyond the possibilities already offered by conventional condensed matter systems. Engineered optical lattices with laser-induced tunnelling amplitudes has enabled the realization of artificial high gauge fields with remarkable tunability.

One of the most interesting developments in ultracold atomic physics is the study of neutral atoms in optical lattices~\cite{IBloch, Greiner, IBloch2}.
On the other hand, it is known that properties of low-dimensional systems may be completely changed by the presence of correlations. Electronic correlation itself can induce several many-body effects like metal-insulator transitions, charge-density waves and antiferromagnetism. Usually, Hubbard model is the prototype correlated system. However, one can also have two types of electrons, considered by Falicov and Kimball, a mobile electron and a massive localized electron, interacting with each other locally. Since the Falicov-Kimball model (FKM) is one of the simplest models of correlated systems and has a number of analytical and numerical results available on square lattices, it can be used to study correlation effects~\cite{FKM1, Freericks, FKM2, FKM3} on Hofstadter spectrum. Moreover, recently, there are proposals for the realization of FKM in optical lattices with mixtures of light atoms and heavy atoms~\cite{Ates, Ziegler} in the context of cold atom systems. A gauge-field can be experimentally realized for ultracold particles (fermions and bosons) in optical lattices~\cite{Dalibard}. Several experiments in optical lattices have confirmed Hofstadter physics\cite{Atala}. Therefore, it is interesting to examine the structure of the Hofstadter butterfly in the presence of electronic correlations.

\begin{figure}[!th]
\includegraphics[trim={3.9cm 0cm 0cm 0cm}, clip, scale=0.24]{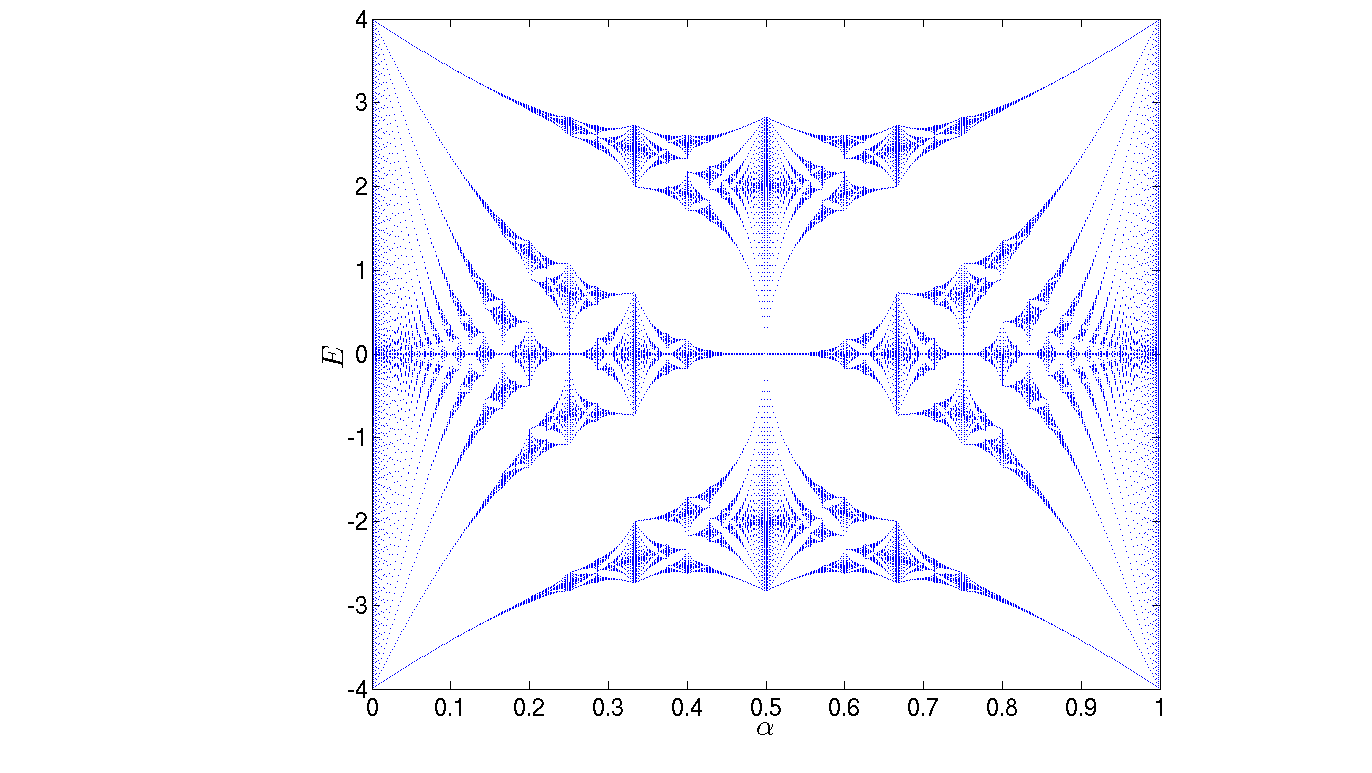}
\caption{\label{Fig.1} (color online)
Energy eigenvalues vs $\alpha$ on an infinite square lattice.} 
\end{figure}

Triggered by the intriguing results on a square lattice, attempts have been made to unravel this structure in a hexagonal lattice to see the effect of lattice geometry on Hofstadter butterfly\cite{Gumbs}. The self-similar fractal energy structure depends sensitively on the geometry of the underlying lattice, as well as the applied magnetic field. Moreover, the finite size of the systems seriously affects the energy spectrum. Electrons on a finite size square lattice in the presence of perpendicular magnetic field were studied in order to see the effects of boundary conditions and Hubbard type of Coulomb correlations recently~\cite{maska}. In this study, we consider a non-bipartite lattice i.e., a triangular lattice in presence of both magnetic field and FKM-type of electron correlation. The particle-hole symmetry of the square lattice is no longer extant. In order to incorporate finite size effects as well as electronic correlations on Hofstadter butterfly, we start with an FKM on a finite system with both open and periodic boundary conditions.

\section{Model}
To study the effect of Coulomb repulsion on Hofstadter spectrum in different lattices we have used a spinless, FKM at half-filling, as we ignore the Zeeman effect which effectively projects out the ``wrong-spin" sector to high energies at high magnetic fields.
\begin{equation}
\begin{aligned}
H_0 =  - {\sum\limits_{ < ij > } {{t}({d_i}} ^\dag }{d_j} + h.c.) + U\sum\limits_i {{d_i}^\dag } {d_i}{f_i}^\dag {f_i} + \hfill \\
{E_f}{\sum\limits_i {{f_i}} ^\dag }{f_i}
\end{aligned}
\end{equation}
\noindent where ${<i,j>}$ are nearest-neighbour site indices on a square lattice (lattice constant $a = 1$), $d_{i} \, (f_{i})$ are itinerant (localized) electron annihilation operators at site $i$. The first term is the kinetic energy due to hopping between nearest neighbours, where, ${t}$ is the hopping integral for $d$-electrons (taken as 1 throughout our calculation). The second term represents on-site Coulomb interaction between $d$ (density ${n_d} = \frac{1}{N}\sum\limits_i {{d_i}^\dagger} {d_i}$) and $f$-electrons (at $E_f$, with density ${n_f} = \frac{1}{N}\sum\limits_i {{f_i}^\dagger} {f_i}$; ${N}$ being the number of sites). This Hamiltonian commutes with $\hat{n}_{f,i}$, in which case local occupancy of $f$-electron is either 0 or 1. Since the $f$-electron occupation number is a `classical variable', the Hamiltonian can be `solved' by using a Monte Carlo annealing over the $f$-electron configurations.

When a magnetic field normal to the plane of the lattice is switched on, the field couples to the spinless, mobile fermions via canonically conjugate momenta only. Zeeman coupling being absent, the field couples to the `orbital degrees'. We choose the Landau gauge $\vec{A}(r)={B(0,ma,0)}$, for a uniform magnetic field of magnitude ${B}$ perpendicular to the 2D-plane of the lattice, electrons propagating along $x$ and $y$ would acquire a different phase shift and that would induce an interference between them, due to the Aharonov-Bohm effect. Thus the $y$-directional hopping picks up a `Peierls phase' ${t_{ij} = -t\exp(\pm ie/\hbar\int_j^i{A(\vec{r})d\vec{r}}})$ = $ {-t\exp (\pm 2\pi im\frac{\phi }{{{\phi _0}}})}$ ; one can consider a plane-wave along this direction. ${\phi}$ = $Ba^{2}$ is the flux per plaquette of a square lattice which is the Aharonov-Bohm (AB) phase around a closed path along the plaquette. We consider only rational magnetic flux, i.e., $\phi = \frac{p}{q} \phi _{0}  = {\alpha}\phi_{0}$ with ${p},\, {q}$ co-prime integers and the Dirac flux quantum is $\phi_0$. Lattice periodicity is lost along the $x$-direction due to the magnetic field. As is customary, the lattice is discretized considering $(x,y) = (ma, nb)$ - each site is then indexed by an integer ``$m$" along x-direction. The Hamiltonian invariant only for lattice translations in the magnetic translation group~\cite{Zak}. This leads to a unit cell $q$ times larger than the original one to accommodate an integer flux $p{\phi_0}$. Therefore, to enfold a magnetic flux $B$ = $\frac{2\pi}{L}$, the magnetic supercell now becomes a strip of length $L$ \cite{polaron} (in this calculation the maximum $L = 24$ is used). In the non-interacting limit $U = 0$, the energy spectrum vs magnetic field plot displays a self-similar structure on a square lattice. As we see from Fig.1, this fractal structure shows some clear gaps, known as Hofstadter gaps which appear and disappear depending on the strength of the  applied magnetic field. 

\begin{figure}[!th]
\includegraphics[trim={5cm 0cm 0cm 0cm}, clip, scale=0.26]{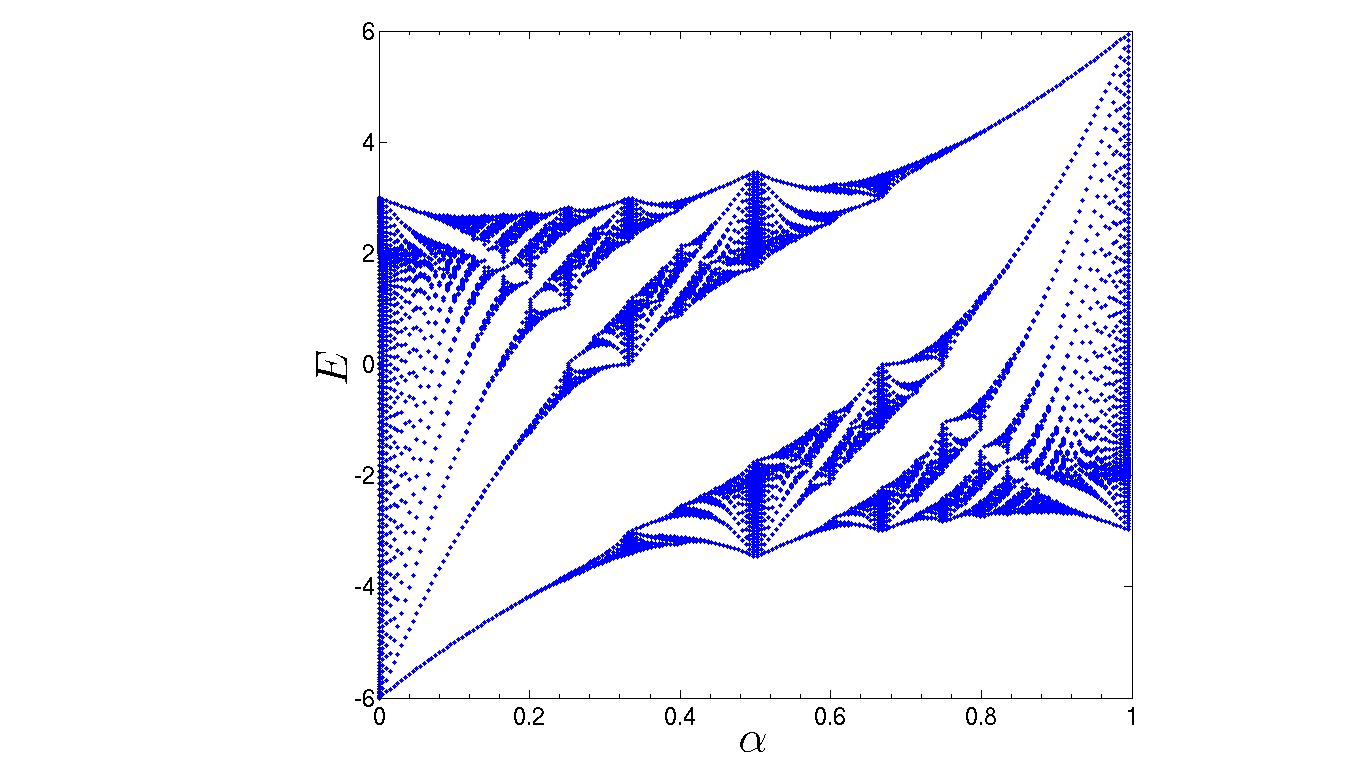}
\caption{\label{Fig.2} (color online)
(a)Hofstadter Butterfly on an infinite triangular lattice for ${t_1 = t_2 = t_3 = 1}$.} 
\end{figure}
Next, we consider the same problem on a triangular lattice. In the absence of a magnetic field, it has a band dispersion $E(k) = 4{t}{{cos\frac{\sqrt{3}k_x{a}}{2}}cos{\frac{k_y{a}}{2}}} + 2tcos{k_y{a}}$ and the atom at the origin has six neighbours at $(\pm{a},0)$ and $({\pm a/2, \pm{\sqrt{3}a/2}}),\, (k_x, k_y)$ represents the wave vector of an electron and $a$ is the lattice constant. With assymetric gauge, we can write down the tight-binding Hamiltonian as below 

\begin{equation}
   \begin{array}{l}
 H = [ - {t_1}\sum\limits_{m,n} {({d^\dag }_{m + 1,n}} {d_{m,n}}\exp (i{\varphi _1}) + h.c.) \\ 
 \,\,\,\,\,\,\,\,\, - {t_2}\sum\limits_{m,n} {({d^\dag }_{m,n + 1}} {d_{m,n}}\exp (i{\varphi _2}) + h.c.) \\ 
 \,\,\,\,\,\,\,\,\, - {t_3}\sum\limits_{m,n} {({d^\dag }_{m,n + 1}} {d_{m + 1,n}}\exp (i{\varphi _3}) + h.c.)]\\{\rm{where}}, \\ 
 {\varphi _{ij}} = 0;\,\,\,\,\,\,\,\,\,\,\,\,\,\,\,\,\,\,\,\,\,\,\,\,\,\,\,\,\,\,\,\,i = (m + 1,n)\,,\,j = (m,n) \\ 
 {\varphi _{ij}} = 2\pi m\phi ;\,\,\,\,\,\,\,\,\,\,\,\,\,\,\,\,\,\,\,\,\,\,i = (m,n)\,\,,j = (m,n + 1) \\ 
 {\varphi _{ij}} = (2\pi m + 1/2)\phi ;\,\,\,\,\,i = (m,n + 1),\,\,\,j = (m + 1,n) \\ 
 \end{array}
\end{equation}
                                                                                                                                                                                                                                                                                                                                                                                                                                                                                                                                                                                                                                                                                                                                                                                                                                                                                                                                                                                                                                                                                                                                                                                                                                                                                                                                                                                                                                                                                                                                                                                                                                                                                                                                                                                                                  If we do a Peierls substitution in the tight-binding Hamiltonian for an electron in a triangular lattice, we will have phases associated with four nearest neighbour hoppings, which makes an angle $\pm {\pi/3}$ and $\pm {{2\pi}/3}$ with respect to the $x$-direction respectively. The hopping parameter along $x$-direction is $t_1$, along ${\pm{\pi/3}}$ direction ${t_2}$, and along $\pm {{2\pi}/3}$-direction ${t_3}$. The lattice is discretized by assuming $(x, y) = (mb, nc)$ where $b = a/2$ and $c = \sqrt{3}a/2$, $\varphi = 2{\sqrt3}Bb^2/{\phi_0}$ is the magnetic flux through each unit cell\cite{Oh}. If we plot the energy eigenvalues with respect to external magnetic flux we get the Hofstadter butterfly for the triangular lattice (see Fig.2)\cite{Claro, Hastugai}.

\section{Finite size effects and electron correlation on Hofstadter spectrum}

\begin{figure}[!th]
\includegraphics[trim={0cm 0cm 0cm 0cm}, clip, scale=0.22]{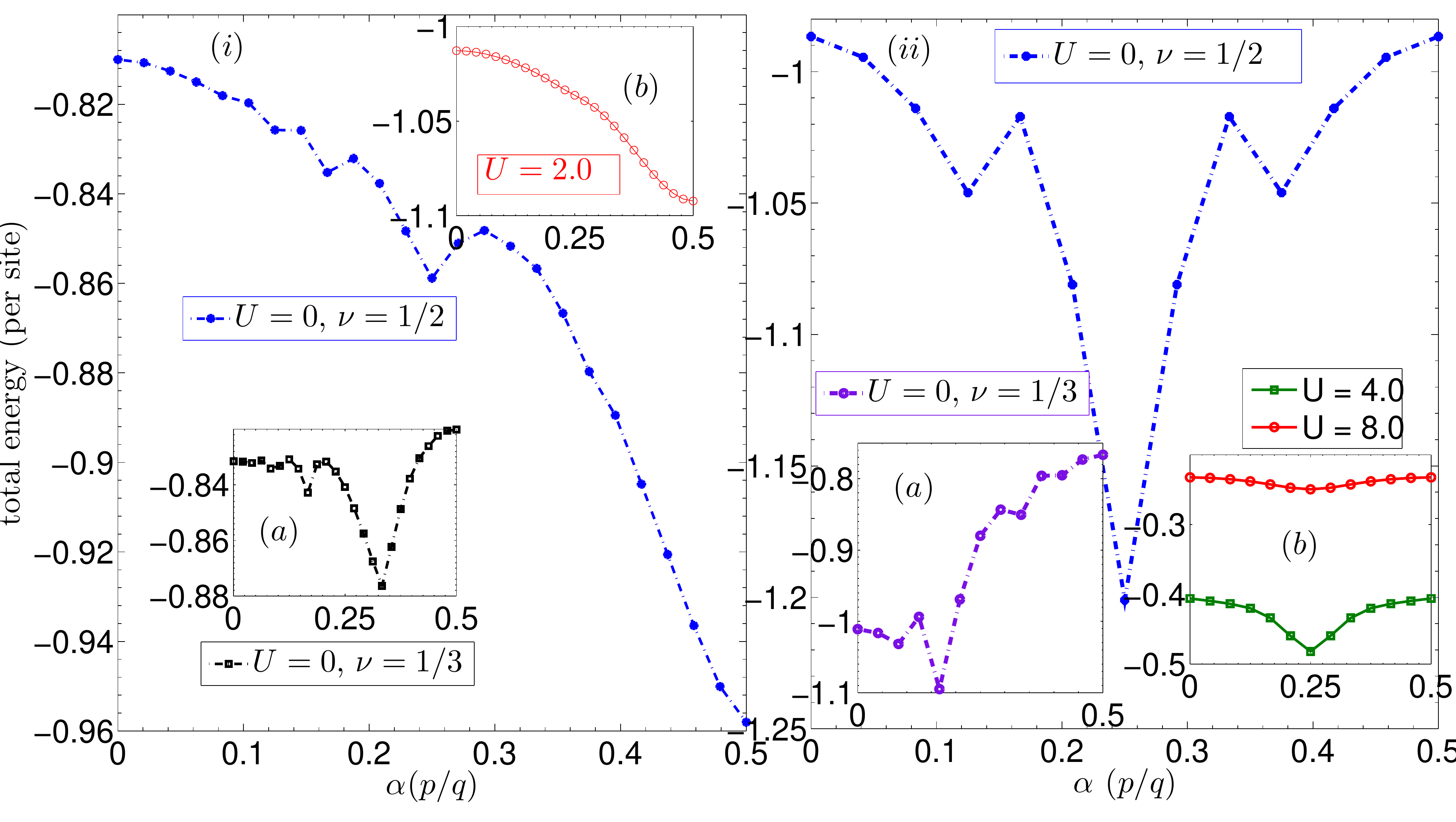}
\caption{\label{Fig.3} (color online)
(i) Total energy per site vs magnetic flux per plaquette for a square lattice at half-filling. Inset (a) for filling $\nu = 1/3$ at $U = 0$ and inset (b) shows total energy variation with $\alpha$ at $\nu = 1/2$ and $U = 2.0$. (ii) Total energy per site vs magnetic flux per plaquette $\alpha$ for a triangular lattice at half-filling. Inset (a) for 1/3 filling and $U = 0$, while inset (b) is total energy vs. $\alpha$ plot for two $U$-values at half-filling.} 
\end{figure}

\begin{figure}[!th]
\includegraphics[trim={3.9cm 0.2cm 2cm 0.5cm}, clip, scale=0.26]{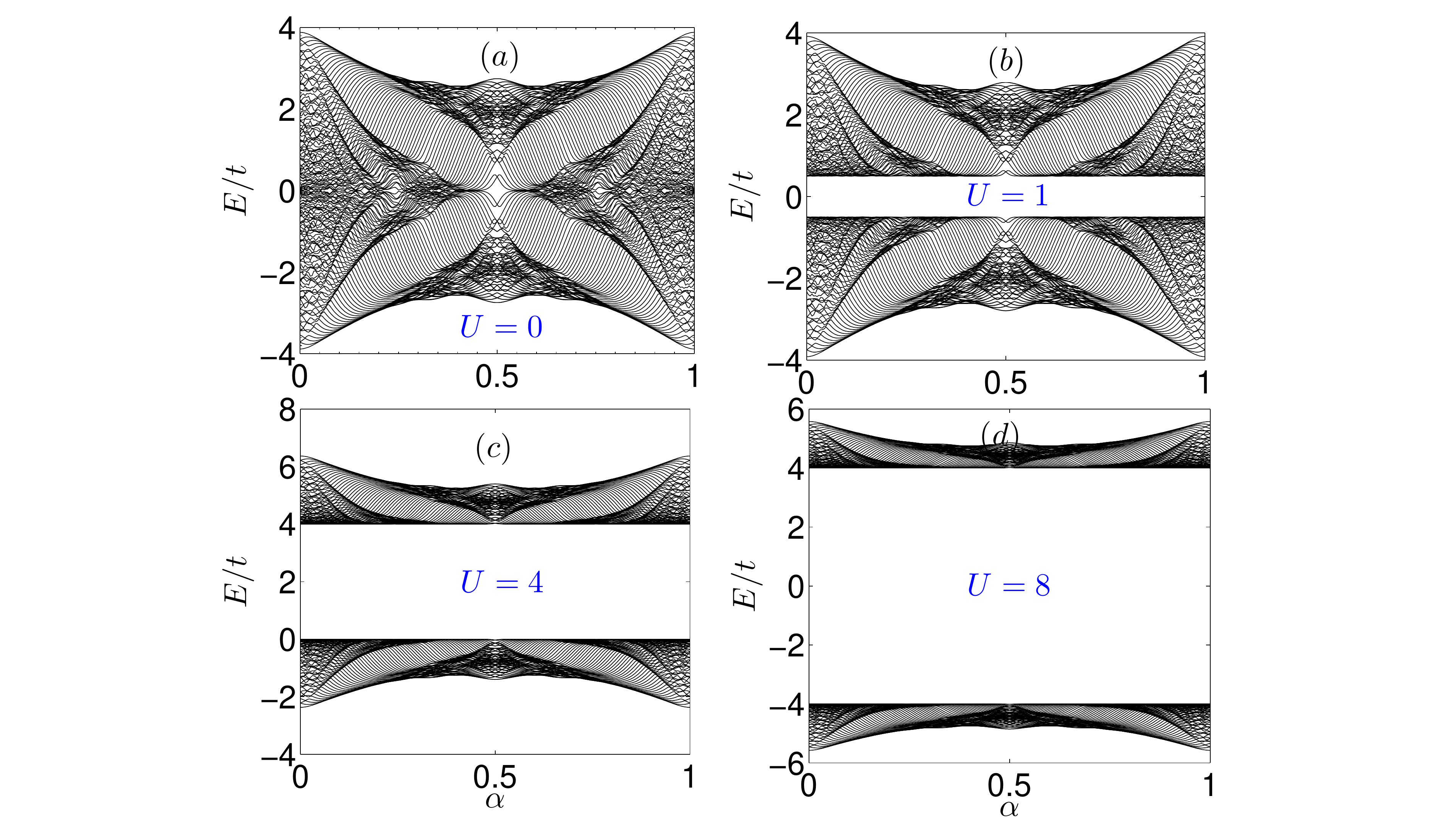}
\caption{\label{Fig.4} (color on-line)
(a) Hofstadter spectrum on a finite size square lattice at different values of $U$.} 
\end{figure}
First, we consider non-interacting two-dimensional systems in presence of a large orbital magnetic field. It was shown by Hasegawa et al.~\cite{Hasegawa} that for tight-binding electrons moving in a strong gauge field in a 2D lattice, there are some special values of flux at which the total energy per site becomes a minimum. For a square lattice, it was found that the energy-minimizing flux $\alpha$ is exactly equal to the electron-filling ($\nu$) whereas for a triangular lattice, it is equal to half of the filling. For a square lattice, at $\nu = \alpha$, the energy is lowered by the lowest group of states just below the largest commensurabilty gap \cite{Hasegawa} at the lower energy side (see Fig.1). This theorem has been checked in our case for both $U = 0$ (see Fig.2, inset 2(a) and Fig.3, inset Fig.3(a)) and $U \ne 0$. We can conclude that this theorem is valid even in the presence of interaction (see Fig.2(b), Fig.3(b)). This rule checks out for all the cases discussed below.
\begin{figure*}[!th]
\includegraphics[trim={4.5cm 0.5cm 0cm 1.5cm}, clip, scale=0.6]{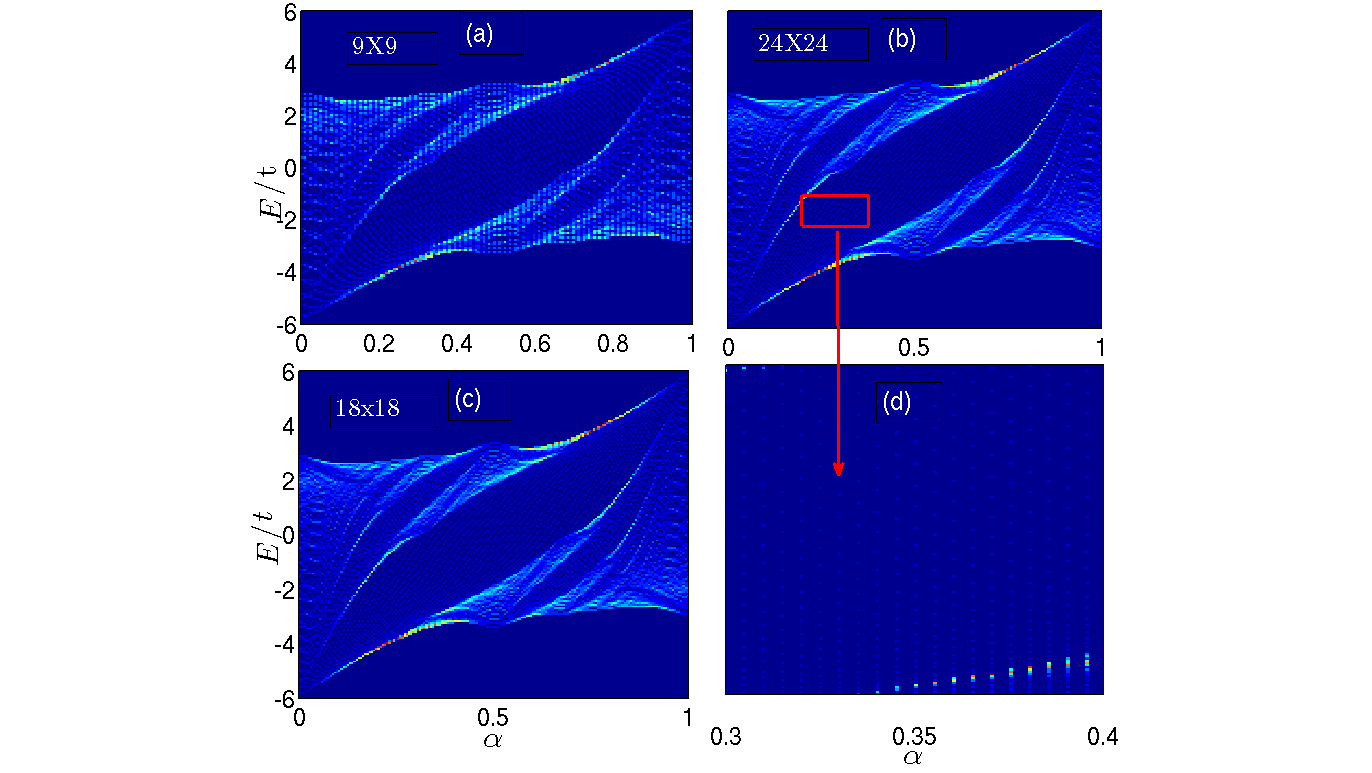}
\caption{\label{Fig.5} (color online)
The energy spectrum vs magnetic flux in a triangular lattice for different system sizes with open boundary condition for $U=0$.} 
\end{figure*}

\begin{figure*}[!th]
\includegraphics[trim={3cm 6cm 8cm 6cm}, clip, scale=0.6]{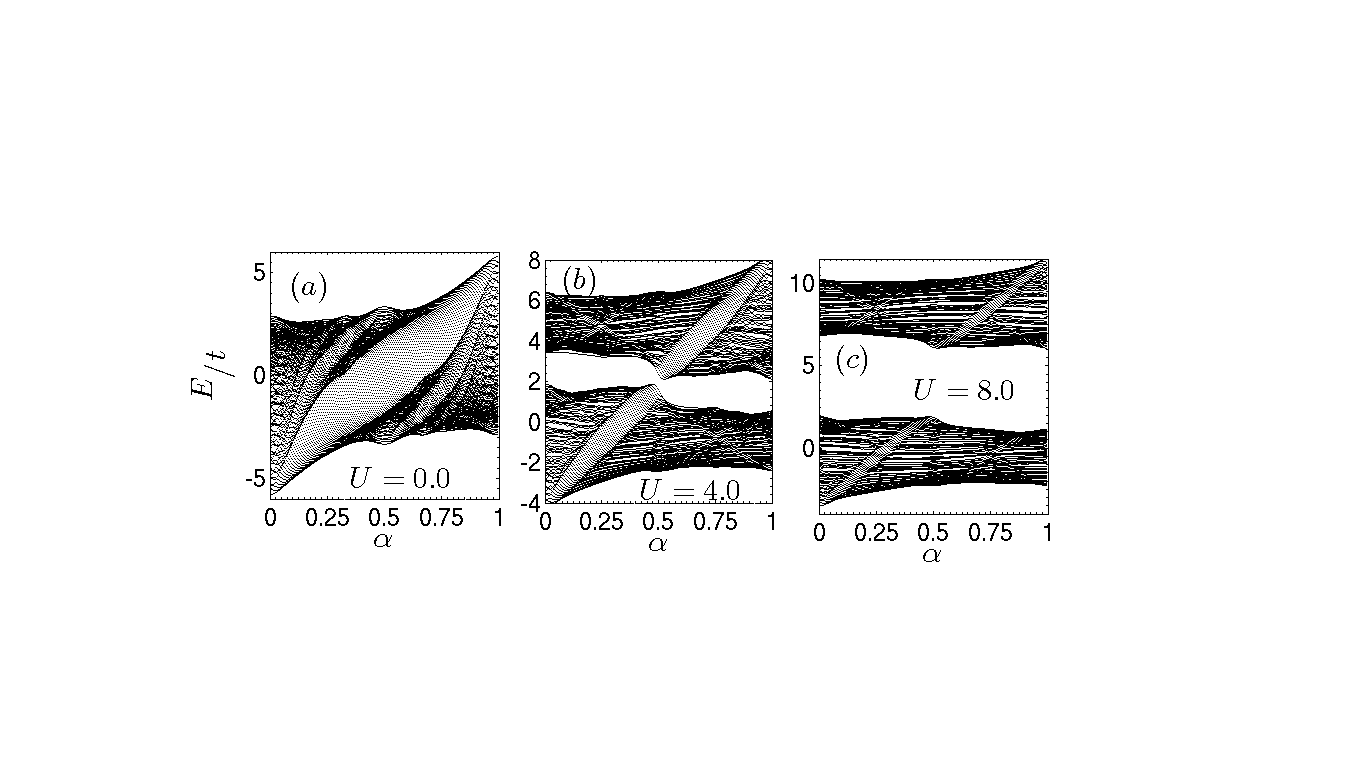}
\caption{\label{Fig.6}
The energy spectrum vs magnetic flux in a finite triangular lattice for different $U$ with open boundary condition.} 
\end{figure*}

\begin{figure*}[!th]
\includegraphics[trim={0cm 2cm 0cm 3.5cm}, clip, scale=0.51]{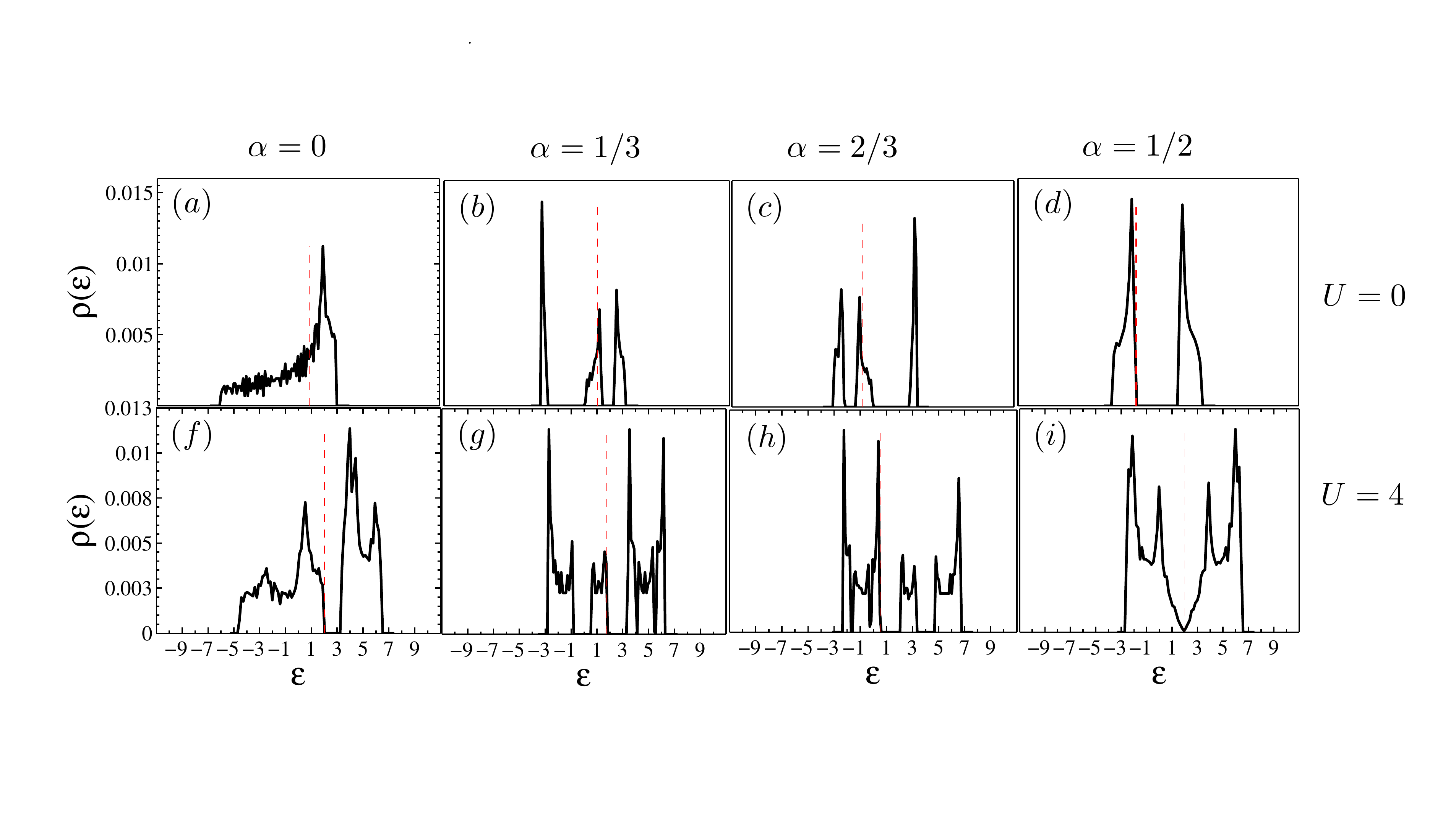}
\caption{\label{Fig.7} (color online)
DOS for triangular lattice with PBC at different values of applied field and electron correlation.} 
\end{figure*}

\begin{figure*}[!th]
\includegraphics[trim={0cm 0cm 1cm 0.15cm}, clip, scale=0.45]{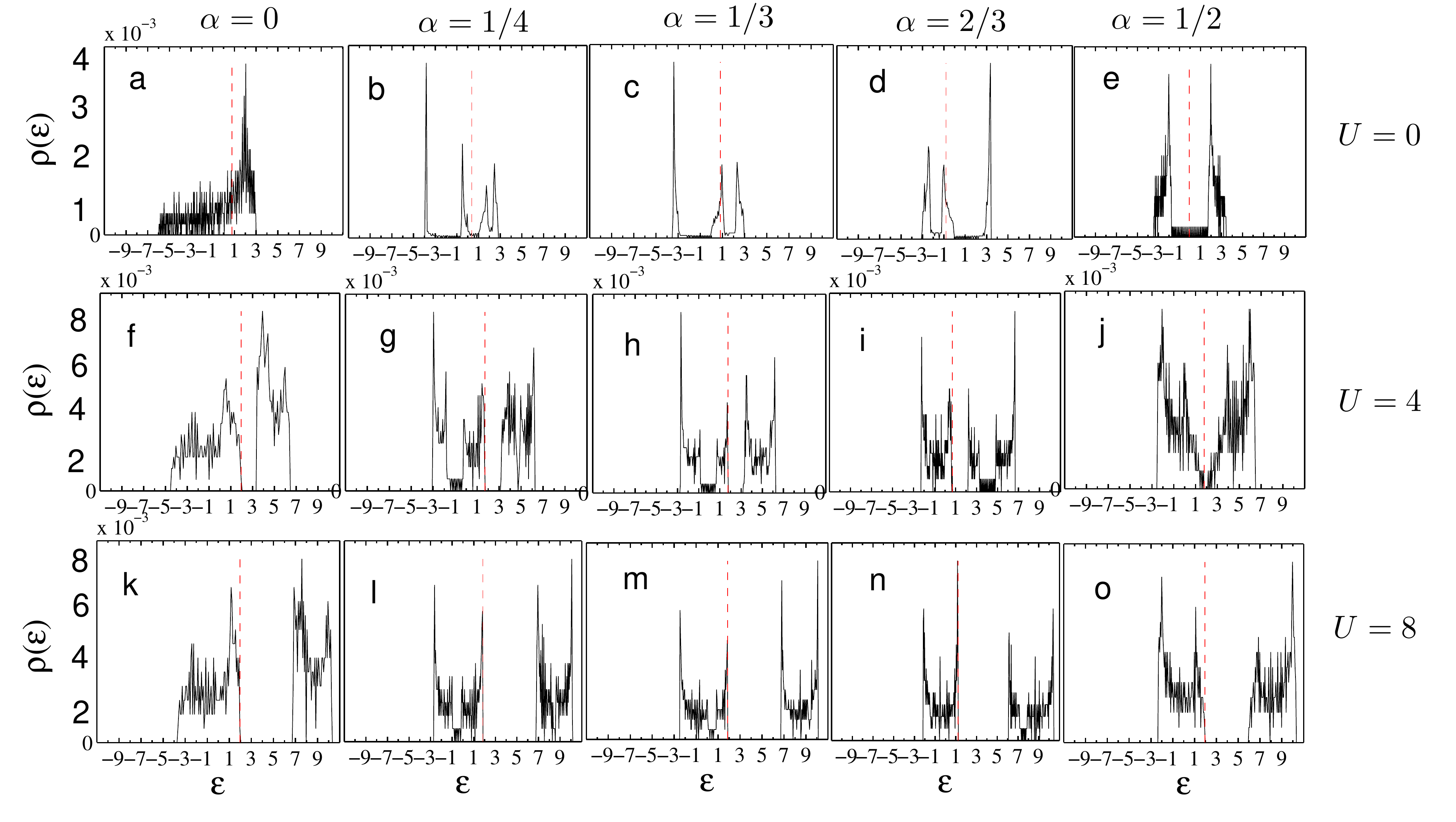}
\caption{\label{Fig.8} (color online)
DOS for a triangular lattice with OBC at different values of applied field and electron correlation.} 
\end{figure*}
We study the effect of a finite boundary on such recursive structures. If a Landau gauge is chosen, we have essentially a plane wave along one direction while the other direction remains open - effectively a cylindrical geometry. For an infinite system, Hofstadter energy spectrum depends precisely on $\alpha$, the magnetic flux per plaquette of the lattice, divided by the unit of flux quantum. For rational $\alpha$ ($=p/q$, $p$ and $q$ are co-prime integers) the spectrum splits into q sub-bands and each state is $q$-fold degenerate. On the other hand, the Hofstadter spectrum with torus geometry i.e., a lattice with periodic boundary conditions and rational magnetic flux through the plaquette, shows well-defined gaps. Although the qualitative aspect of the spectrum remains invariant  for several kinds of boundaries, some extra states show up inside the Hofstadter gaps. Fig.4(a) shows these extra states appearing inside the large and small gaps in the Hofstadter spectrum for a finite size square lattice. These states remain even if the system size is increased. We introduce the FKM-type Coulomb correlation on this structure at half-filling, the half-filled FKM has a charge-density-wave (CDW) ground state \cite{Brandt,Mielsch}. At half-filling the localized $f$-electrons fill up one of the two sublattices of the square lattice (the so called chessboard structure) and the corresponding ground state is insulating for all $U \ge 0$. At half-filling, as we increase $U$, the gap also increases and the gap in the spectrum has exactly the same value as $U$.

We examine the effects of lattice sizes on the energy spectrum for a triangular lattice. As seen from Fig.5 the spectrum contains some extra states (compare with Fig.2) which survive even if we increase the system sizes from $9 \times 9$ to $24 \times 24$. Then we study the effect of Coulomb correlations on these states at different magnetic fluxes. In the FKM, the $d$-electrons sample the annealed disordered background of $f$-electrons ($n_{f,i}=0,1$ only) and a Monte Carlo annealing over all $f$-electron configurations get us the minimum energy configuration. It was found that at half-filling the triangular lattice has a diagonal stripe pattern in the range of $U = 4$ to $10$, which confirms earlier findings. For lower values of $U$, the ground state does not show any specific order \cite{Yadav}. We have obtained the ordered ground state configuration at half-filling and used it for our calculations. We compute the density of states (DOS) of $d$-electrons for various values of $U$ on a $24\times24$ lattice under both periodic (PBC) and open boundary conditions (OBC). The structure of the Hofstadter butterfly is greatly modified and Hofstadter band-gaps and bandwidths are also modified under the effect of Coulomb repulsion (see Fig.6, also compare with Fig.2). We have shown in Fig.8 the variation in the DOS at different magnetic fields for several $U$ values. 

Figs.7 and 8 show how the DOS depends on the magnetic field and the magnitude of the Coulomb repulsion simultaneously. There are theoretical reports of localization induced by random magnetic fields in a 2D noninteracting electronic system, confirmed by several groups \cite{Akira}. In the context of 2D localization, magnetic fields can induce a metal-insulator transition by controlling the degeneracy of the Landau levels \cite{abrahams}. In our case, even in the absence of electron correlation one can see gap induced in the DOS (at $\alpha=0.5$) solely due to the external magnetic field. This is an example, where one can see metal to insulator transition due to application of external magnetic field. For weak to moderate Coulomb repulsion, the DOS strongly depends on the magnetic field. On the other hand, in the absence of a magnetic field, electronic correlations can also open up a gap in the DOS (Fig.8(f)). For a triangular lattice with open boundary conditions, the field brings in extra states inside the Hofstadter gaps, which appear at rational values of flux. For a strong magnetic field the DOS becomes robust against Coulomb correlation. This result already suggests that the Coulomb interaction modifies various parts of the Hofstadter butterfly in different ways. As seen from the plot, states which appear due to the finite size effects progressively vanish with the addition of Coulomb repulsion for some values of the flux within the range of $U$ used here. A further increase of $U$ causes an enhanced band gap. Although not shown here, at larger values of $U$ no further qualitative change in the DOS occurs, except a larger band gap and narrower bandwidth of the spectrum. We calculate the magnitude of the gap with the magnetic field as well as $U$. As it is seen from Fig.9, the gap varies in a different way in the presence and absence of correlation. In the non-interacting limit, the gap has a maximum at $\alpha = 0.5$, and the opposite happens at the same place for finite $U$. 
\begin{figure}[!th]
\includegraphics[trim={1cm 0.5cm 0.5cm 0cm}, clip, scale=0.21]{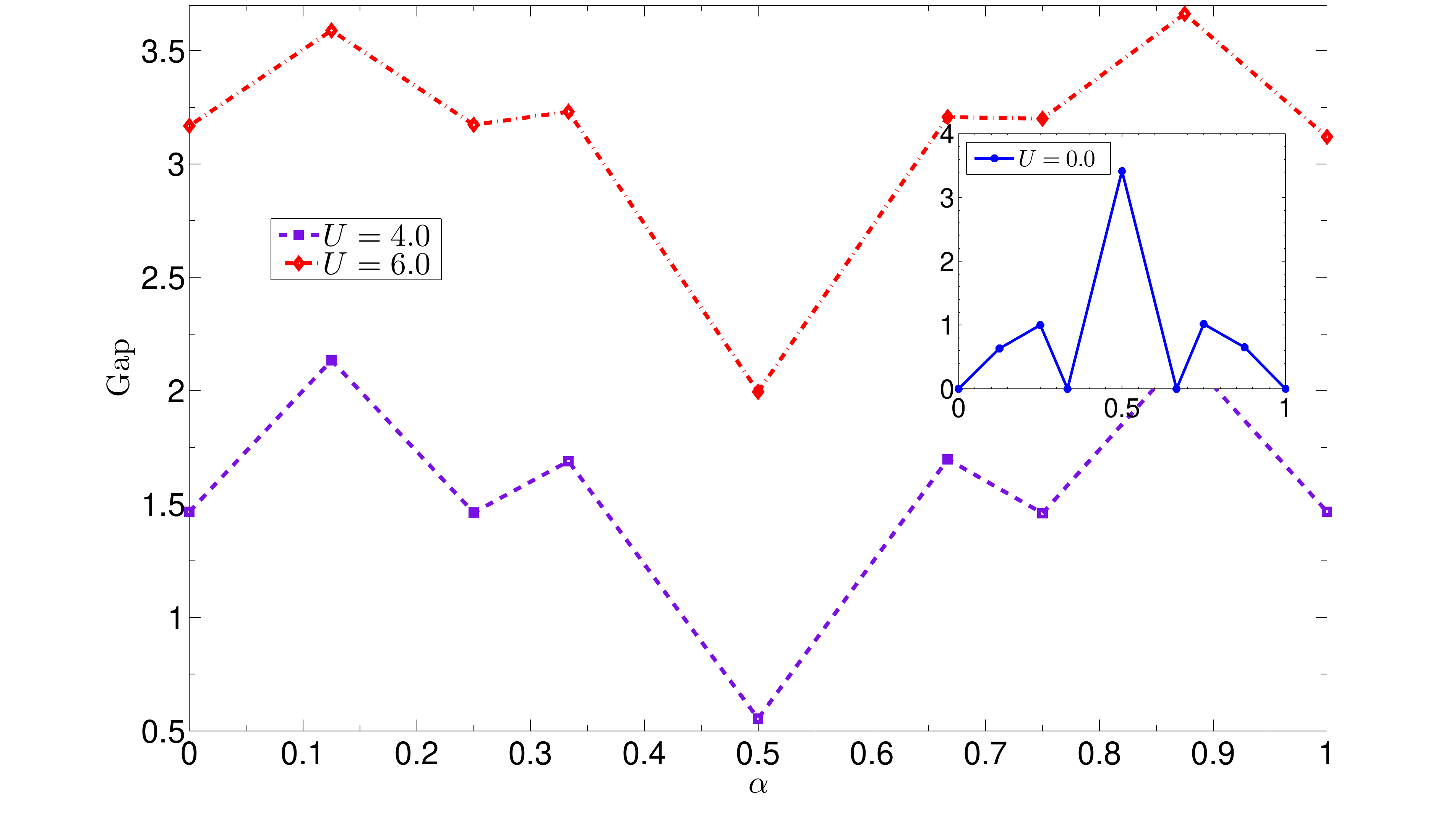}
\caption{\label{Fig.9} (color online)
Magnitude of the gap in DOS as a function of $U$ for various magnetic fields in a triangular lattice. Inset shows the gaps for the noninteracting limit. These results are for a $24\times24$ lattice with pbc.} 
\end{figure}

\section {Orbital Current}

Introduction of the magnetic field in the hopping term of the Hamiltonian of Eq(1) breaks the temporal invariance of the Hamiltonian and gives rise to persistent current of the charged particles. For any bond connecting neighboring sites $(i,j) = (n,m; n, m \pm 1)$ of the lattice, we define $t_{ij} = -t$ for hopping along $x$; ${t_{ij} = -t\exp(\pm ie/\hbar\int_j^i{A(\vec{r})d\vec{r}}})$ = $ {-t\exp (\pm 2\pi im\frac{\phi }{{{\phi _0}}})}$ for hopping along $y$-direction. We calculate the local bond current $v_{ij}=\frac {1}{i\hbar}[t_{ij}d_{i}^\dagger d_{j} - t_{ij}^*d_{j}^\dagger d_{i}]$,

\begin{figure}[!th]
\includegraphics[trim={3.1cm 5.0cm 3.0cm 5.0cm}, clip, scale=0.59]{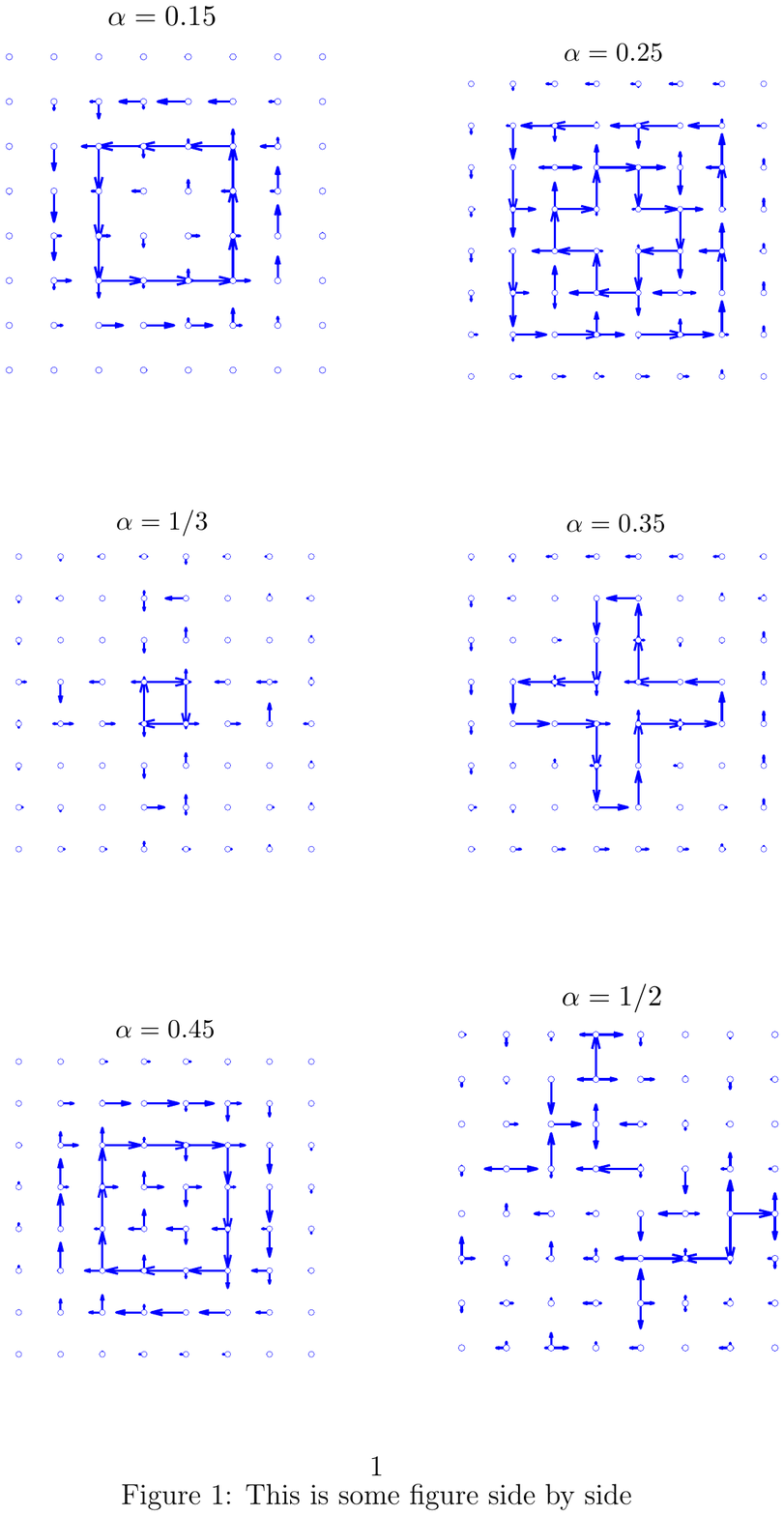}
\caption{\label{Fig.10} (color online)
Current patterns on a finite square lattice at half-filling for different flux.} 
\end{figure}

\begin{figure*}[!th]
\includegraphics[trim={3.7cm 20cm 0cm 4.5cm}, clip, scale= 1.15]{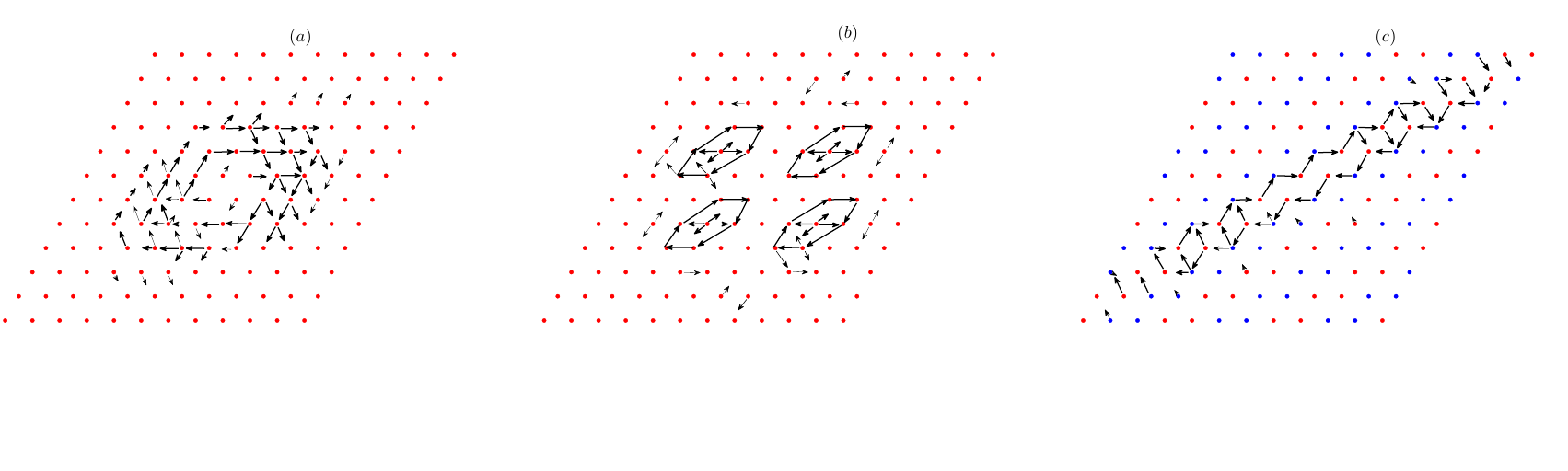}
\caption{\label{Fig.11} (color online)
Current patterns on a finite triangular lattice at half-filling for (a) $\alpha = 0.15$, (b) $\alpha = 1/3$ at $U = 0$. (c) shows current for $\alpha = 2/3$ at $U = 10$..}
\end{figure*}

\begin{figure}[!th]
\includegraphics[trim={2cm 0cm 0cm 0cm}, clip, scale=0.23]{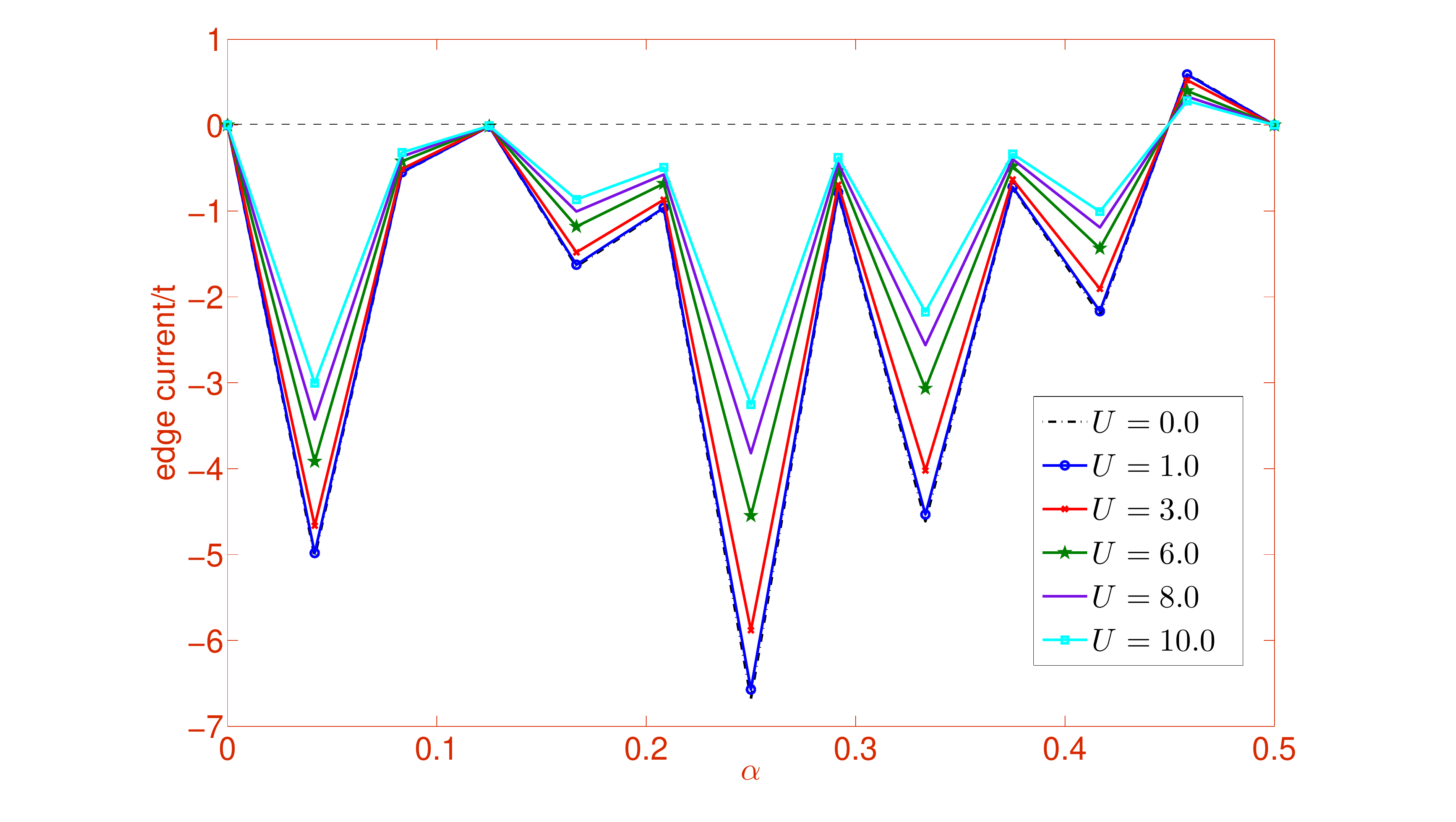}
\caption{\label{Fig.12} (color online)
Edge current for a half-filled square lattice for different $U$.} 
\end{figure}

\begin{figure}[!th]
\includegraphics[trim={2cm 0cm 0cm 0cm}, clip, scale=0.23]{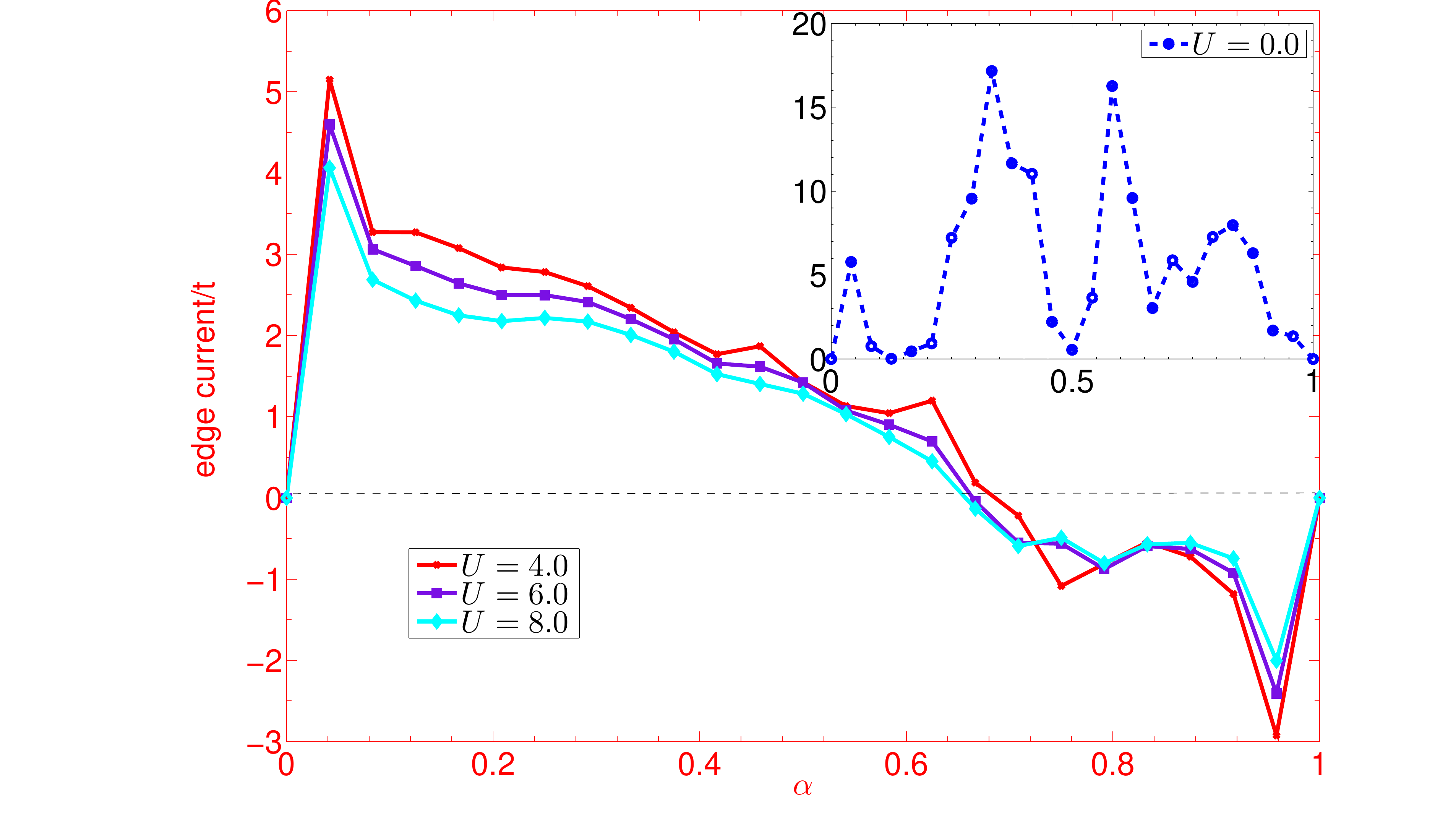}
\caption{\label{Fig.13} (color online)
Edge current on a triangular lattice at half-filling for different values of $U$; inset for $U=0$.} 
\end{figure}

The results obtained from exact diagonalization for an $8\times8$ square lattice are summarized. Note that for a square lattice we restrict ourselves to the interval $\alpha$  ($0 - 0.5$), as the Hamiltonian is invariant under $\alpha \rightarrow 1 - \alpha $, while $\alpha \rightarrow - \alpha$ only changes the magnetic field direction. Fig.10 shows the current profile in the ground state for different interaction strengths and $\alpha$ values for a square lattice. The current patterns found for all $U$ look similar to those obtained at $U = 0$ although the magnitude of current reduces with $U$. Note that the central current reverses sign beyond some critical flux $\alpha = 0.43$, and this value is same for all $U$. The edge current shows an oscillation with $\alpha$, in the presence of $U$: a reduction in the magnitude of edge current is observed, but the pattern of oscillation is fixed in all cases. As there is a change of sign of orbital current at about $\alpha = 0.43$, the same is reflected for the boundary current; at this flux, edge current goes to zero and reverses sign. There is a similar study on the currents on a finite square lattice in a Bose Hubbard model \cite{current}. The current rotates in opposite direction at some specific range of flux in that case as well. Since the effect exists even when $U=0$ and in the bulk too, it is connected to the change in the Fermi surface topology (particle-like or hole-like) at that flux as seen in the Hofstadter spectrum at about $\phi=0.43$. 

The same calculation has been done on a triangular lattice for which $ E(\alpha) = -E(1 - \alpha)$ with $0 \le \alpha \le 1/2$ at half-filling. Fig. 11 shows current configurations for some magnetic flux for a half-filled triangular lattice in the non-interacting as well as finite-$U$ regime. As $U$ is increased, current is found to avoid the sites which are occupied by $f$-electrons and flow though a narrow channel (Fig. 11(c)). Edge current in an interacting triangular lattice at half-filling changes sign at $\alpha = 2/3$, although there is no such regular pattern in the edge current variation at a finite $U$. 

\section{Summary and conclusion}
We have checked for the finite size effects on a triangular lattice in the presence of magnetic field. Further, we study the effect of Coulomb interaction on this system. There is  a competition between the applied magnetic field and the electronic correlation. The magnetic field induces a gap (largest at $\alpha=0.5$) in the density of states even in the absence of correlation. At finite $U$, however, this gap reduces from its $U=0$ value and becomes smallest at $\alpha=0.5$. For a finite size system, the Hofstadter gaps are filled up with some extra states and these states go away with the inclusion of correlation in some cases. 

\acknowledgements
The author acknowledges useful discussions with A. Taraphder, Monodeep Chakraborty, Diptiman Sen and S. Lal. Centre for Theoretical Studies, IIT Kharagpur is acknowledged for providing computer facilities.

\end{document}